\documentclass[preprint,showpacs,preprintnumbers,pre,superscriptaddress]{revtex4-1}
\usepackage{graphicx}
\usepackage{amssymb}
\usepackage{amsmath,bm}
\usepackage{amsfonts,color}
\usepackage[T1]{fontenc}
\begin{document}

\title{Equilibrium free energy differences from a linear nonequilibrium equality}
\author{Geng Li}
\affiliation{CAS Key Laboratory for Theoretical Physics, Institute of Theoretical Physics, Chinese Academy of Sciences, Beijing 100190, China}
\affiliation{Graduate School of China Academy of Engineering Physics, Beijing 100193, China}
\author{Z. C. Tu}\email[Corresponding author. Email: ]{tuzc@bnu.edu.cn}
\affiliation{Department of Physics, Beijing Normal University, Beijing 100875, China}

\begin{abstract}
Extracting equilibrium information from nonequilibrium measurements is a challenge task of great importance in understanding the thermodynamic properties of physical, chemical, and biological systems. The discovery of the Jarzynski equality illumines the way to estimate the equilibrium free energy difference from the work performed in nonequilibrium driving processes. However, the nonlinear (exponential) relation causes the poor convergence of the Jarzynski equality. Here, we propose a concise method to estimate the free energy difference through a linear nonequilibrium equality which inherently converges faster than nonlinear nonequilibrium equalities. This linear nonequilibrium equality relies on an accelerated isothermal process which is realized by using a unified variational approach, named variational shortcuts to isothermality. We apply our method to an underdamped Brownian particle moving in a double-well potential. The simulations confirm that the method can be used to accurately estimate the free energy difference with high efficiency. Especially during fast driving processes with high dissipation, the method can improve the accuracy by more than an order of magnitude compared with the estimator based on the nonlinear nonequilibrium equality.

\end{abstract}
\maketitle

\section{Introduction}

How can one extract equilibrium information from nonequilibrium measurements? This may appear a contradictory question at first glance. According to the second law of thermodynamics, the mean work performed in a nonequilibrium driving process will be larger than the the free energy difference between equilibrium states. Only if a system slowly evolves along a succession of equilibrium states, for example in an isothermal process, will the mean work equal the free energy difference. Recent advances in nonequilibrium statistical mechanics have suggested a promising direction to extract free energy information from nonequilibrium measurements. As one of the most representative achievements, the Jarzynski equality~\cite{Jarzynski1997} establishes a rigorous relation between the free energy difference and the exponential average over the work performed in a nonequilibrium driving process, thus extending the inequality of the second law of thermodynamics. While the equality implies that one can estimate the free energy difference by using arbitrary fast measurements, its applicability is hampered by the poor convergence that arises from the sensitivity of the nonlinear (exponential) average to rare events~\cite{Zuckerman2002,Gore2003,Jarzynski2006}. An illuminating question is whether we can find a linear nonequilibrium equality to avoid this shortage, so that we can pave the way for efficiently extracting equilibrium information from nonequilibrium measurements.

Recently, the present authors and their coworker proposed a concept of shortcuts to isothermality and found a linear nonequilibrium equality (see Eq.~(\ref{eq:meaniw}) below, named the intrinsic work equality) between the free energy difference and the intrinsic work~\cite{Geng2017}, which enlighten us a possible solution to the above question. As a unified framework to accelerate the isothermal process, shortcuts to isothermality have been successfully validated in experiment~\cite{Albay2019,Albay2020A,Albay2020N} and further extended to the optimization of finite-time heat engines~\cite{Pancotti2019,Nakamura2020,Plata2020} and the control of biological evolutions~\cite{Iram2019}. A key point in shortcuts to isothermality is to apply an auxiliary potential to the system of interest, such that the system evolves along the ``isothermal'' line corresponding to the original Hamiltonian. It is this ``isothermality'' that leads to the intrinsic work equality. This equality may be used to estimate the free energy difference with high accuracy, so it provides a new scheme for the free energy estimation in complex systems. Unfortunately, there is an obstacle that the process of solving the auxiliary potential requires the free energy information in advance. Similar obstacles are prevalent in many other schemes for estimating the free energy difference~\cite{Miller2000,Jarzynski2002,Vaikuntanathan2008,Minh2011}. It will be of great significance to the free energy estimation if we could find a method to calculate the auxiliary potential without resort to the free energy information.

In this work, we overcome this obstacle by developing a unified variational approach to approximately realize shortcuts to isothermality. Relying on the accelerated isothermality, we can estimate the free energy difference by using the linear nonequilibrium equality. As a specific application, we consider an underdamped Brownian particle moving in a double-well potential, for which we show that our method improves the accuracy by more than an order of magnitude and shows excellent convergence in fast driving processes. Therefore, our method offers a possible solution to the difficulties of high-efficiency free energy estimation in complex systems, such as biological or chemical molecules.

\section{Shortcuts to Isothermality\label{defwh}}
Shortcut to isothermality~\cite{Geng2017} is a unified framework to accelerate the conventional isothermal process and thereby realize finite-rate transitions between two equilibrium states at the same temperature. In the following, we briefly introduce the strategy of shortcuts to isothermality. Consider a system described by the Hamiltonian $H_{o}(\bm{x},\lambda(t))$ with $\bm{x}=(x_{1}, x_{2}, \cdots,x_{N})$ representing the microstate of the system and $\lambda(t)$ being an externally controlling parameter. The system is coupled to a thermal reservoir with a constant temperature $T$. The motion of the system is governed by the following equation
\begin{equation} \dot{ x }_{i}=f_{i}^{o}(\bm{x},t). \label{eq:motioneq}\end{equation}
In this work, the dot above a variable represents the time derivative of that variable. $\bm{f}^{o}=(f_{1}^{o},f_{2}^{o},\cdots,f_{N}^{o})$ represents a generalized ``force'' field that depends on the Hamiltonian $H_{o}(\bm{x},\lambda(t))$ and the specific dynamics we are considering.

We introduce an auxiliary potential $U_{a}(\bm{x},t)$ to the original Hamiltonian $H_{o}(\bm{x},\lambda(t))$ so that the system distribution $\rho(\bm{x},t)$ is always in the instantaneous canonical distribution of the original Hamiltonian
\begin{equation} \rho(\bm{x},t)= \mathrm{e}^{\beta[F(\lambda(t))-H_{o}(\bm{x},\lambda(t))]}, \label{eq:ieqdistri}\end{equation}
where $\beta=1/k_{B}T$ with $k_{B}$ being the Boltzmann factor.
\begin{equation} F(\lambda)\equiv -\beta^{-1}\ln \left [  \int d \bm{x}\mathrm{e}^{-\beta H_{o}(\bm{x},\lambda)}   \right] \label{eq:ieqfe}\end{equation}
denotes the free energy of the original system in equilibrium for fixed $\lambda$. With additional requirements that $U_{a}(\bm{x},t)$ vanishes at two endpoints of the driving process, the system of interest will appear to evolve along the isothermal line in a finite rate. Along this ``isothermal'' line, we can derive an equality between the free energy difference and the mean work related to the original Hamiltonian (which is called the intrinsic work)~\cite{Geng2017}:
\begin{equation} \Delta F = \langle w_{i} \rangle \equiv\int_{0}^{\tau} \left \langle\frac{\partial H_{o}}{\partial t} \right \rangle dt,  \label{eq:meaniw}\end{equation}
where $\langle \cdots \rangle$ denotes the ensemble average over trajectories. Since Eq.~(\ref{eq:meaniw}) takes a linear average over the work, it inherently converges faster than other nonlinear nonequilibrium equalities~\cite{Zuckerman2002}. In addition to shortcuts to isothermality, many researchers have also discussed the realization of finite-rate transitions from an equilibrium state to another one with the same temperature. One of the most important achievements is the engineered swift equilibration~\cite{Martinez2016NP} proposed by Mart\'{i}nez and coworkers. They realized fast switches between equilibrium states of a Brownian particle system for the first time, see also~\cite{Cunuder2016,Chupeau2018}. Since the instantaneous canonical state~(\ref{eq:ieqdistri}) is not always guaranteed in the engineered swift equilibration, the intrinsic work equality~(\ref{eq:meaniw}) can not be derived from this protocol.

Within the framework of shortcuts to isothermality, the motion equation~(\ref{eq:motioneq}) is modified to the form:
\begin{equation} \dot{x}_{i} = f_{i}^{o}(\bm{x},t)+ f_{i}^{a}(\bm{x},t), \label{eq:modifymoeq}\end{equation}
with $\bm{f}^{a}=(f^{a}_{1},f^{a}_{2},\cdots,f^{a}_{N})$ representing the auxiliary field induced by $U_{a}(\bm{x},t)$. The form of $\bm{f}^{a}(\bm{x},t)$ also depends on the specific dynamics we are considering. The evolution equation of the system distribution $\rho(\bm{x},t)$ can be formally written as
\begin{equation}  \frac{\partial \rho}{\partial t} = \hat{L}_{o} \rho - \frac{\partial }{\partial x_{i}}(f^{a}_{i}\rho ), \label{eq:observG1}\end{equation}
where $\hat{L}_{o}$ represents the evolution operator related to the original field $\bm{ f}^{o}(\bm{x},t)$.
Throughout this paper, the repeated subscripts abide by the Einstein summation convention. We assume that when $\lambda$ is fixed, the original system will relax toward a unique equilibrium state $\rho^{\mathrm{eq}} \propto \mathrm{e}^{-\beta H_{o}(\bm{x},\lambda)}$. Hence, we can obtain $\hat{L}_{o}    \mathrm{e}^{-\beta H_{o}(\bm{x},\lambda)}=0   $. Substituting the instantaneous canonical distribution~(\ref{eq:ieqdistri}) into the evolution Eq.~(\ref{eq:observG1}), we can derive that (see Appendix~\ref{SSec-one} for details)
\begin{equation} f_{i}^{a} \frac{\partial H_{o}}{\partial x_{i}} - \frac{1}{\beta} \frac{\partial f_{i}^{a}}{\partial x_{i}} =  \frac{d F}{d t} - \frac{\partial H_{o}}{\partial t}. \label{eq:G1evoequ}\end{equation}
Similar equation was also derived by Vaikuntanathan and Jarzynski~\cite{Vaikuntanathan2008}. They did not provide a general strategy to solve for the auxiliary field $\bm{f}^{a}(\bm{x},t)$, but suggested to guess the auxiliary field according to physical insight, experience, and prior knowledge of the system.

Equation~(\ref{eq:G1evoequ}) highlights the difficulty of finding the auxiliary field $\bm{f}^{a}(\bm{x},t)$~(or the auxiliary potential~$U_{a}(\bm{x},t)$) precisely: before solving the equation, we need to know in advance the time-dependence of the free energy which is usually hard to obtain for most complex systems. Thus, our goal is to propose a variational method that allows one to circumvent the requirement relating to the free energy information and determine the best possible $\bm{f}^{a}(\bm{x},t)$ under some restrictions, such as some specific boundary conditions or just experimental feasibility.

\section{Variational Shortcuts to Isothermality\label{enproft}}
Based on Eq.~(\ref{eq:G1evoequ}), we can define a function
\begin{equation} \mathcal{D}(\bm{f}) \equiv   f_{i} \frac{\partial H_{o}}{\partial x_{i}} - \frac{1}{\beta} \frac{\partial f_{i}}{\partial x_{i}} +  \frac{\partial H_{o}}{\partial t}  -  \frac{d F}{d t},    \label{eq:functionalG1}\end{equation}
where $\bm{f}=(f_{1},f_{2},\cdots,f_{N})$ represents an approximation to the exact auxiliary field $\bm{f}^{a}(\bm{x},t)$. If $\bm{f}=\bm{f}^{a}$, then $\mathcal{D}(\bm{f})=0$.

The Gauss principle of least constraint~\cite{Gauss1829} provides a clue to seek the best possible $\bm{f}^{a}(\bm{x},t)$. The Gauss principle states that the difference between the trajectory of a restricted system $a_{i}^{r}$ and its unrestricted Newtonian counterpart $a_{i}^{u}=F_{i}/m_{i}$ can be evaluated by the least value of the so called ``constraint'' (Zwang) $Z=m_{i}(a_{i}^{r}-a_{i}^{u})^{2}$. Here, $a_{i}^{r}$ represents the acceleration of the restricted motion, while $a_{i}^{u}$ represents the acceleration of the free unrestricted motion which is defined by force $F_{i}$ divided by mass $m_{i}$. Various efforts have been made to extend the idea of the Gauss principle to other similar problems, such as the development of time reversible deterministic thermostats~\cite{Evans1983,Bright2005} and local quantum counterdiabatic driving protocols~\cite{Sels2017,Kolodrubetz2017,Claeys2019}. Despite the fundamental status of the Gauss principle, there are few reports about the extension of the principle in nonequilibrium driving processes.

Enlightened by the Gauss principle of least constraint, we can define a functional

\begin{equation}  \mathcal{G} (\bm{f}) \equiv  \int d\bm{x} \mathcal{D}^{2}(\bm{f}) \mathrm{e}^{-\beta H_{o}},  \label{eq:Saction}\end{equation}
as a nonequilibrium ``constraint'' on the approximate auxiliary field $\bm{f}(\bm{x},t)$. Here we have multiplied the local constraint $\mathcal{D}^{2}(\bm{f})$ by a function $\mathrm{e}^{-\beta H_{o}}$ and then taken an integral over the whole phase space. In principle, the function $\mathrm{e}^{-\beta H_{o}}$ can be replaced by any positive function. We will find that the function $\mathrm{e}^{-\beta H_{o}}$ can help eliminate the free energy information in the nonequilibrium constraint~(\ref{eq:Saction}). We can prove that finding the exact auxiliary field in Eq.~(\ref{eq:G1evoequ}) is equivalent to solving the variational equation (see Appendix~\ref{SSec-two})
\begin{equation} \frac{\delta \mathcal{G}(\bm{f})}{\delta \bm{f} } =0.  \label{eq:ELequ}\end{equation}

Substituting Eq.~(\ref{eq:functionalG1}) into the nonequilibrium constraint~(\ref{eq:Saction}), we can derive
\begin{align}  \mathcal{G} (\bm{f}) = & \int d \bm{x} \left(     f_{i} \frac{\partial H_{o}}{\partial x_{i}} - \frac{1}{\beta} \frac{\partial f_{i}}{\partial x_{i}} \right)^{2}\mathrm{e}^{-\beta H_{o}} \nonumber \\ &  -\frac{2}{\beta} \int d \bm{x} \left( \frac{\partial H_{o}}{\partial t}  -  \frac{d F}{d t}     \right) \frac{\partial}{\partial x_{i}} (f_{i} \mathrm{e}^{-\beta H_{o}}) \nonumber \\ &+ \int d \bm{x} \left( \frac{\partial H_{o}}{\partial t} -  \frac{d F}{d t}           \right)^{2}\mathrm{e}^{-\beta H_{o}} , \label{eq:Sabefore}\end{align}
which reveals that the nonequilibrium constraint is closely related to the time derivative of the system free energy, $d F/d t$. The third term of Eq.~(\ref{eq:Sabefore}) does not affect the variation in Eq.~(\ref{eq:ELequ}) since it is independent of $\bm{f}(\bm{x},t)$. By using integration by parts, we can eliminate $d F/d t$ in the second term of the constraint~(\ref{eq:Sabefore}):
\begin{align}   &-\frac{2}{ \beta} \int d \bm{x} \left( \frac{\partial H_{o}}{\partial t}  -  \frac{d F}{d t}     \right) \frac{\partial}{\partial x_{i}} (f_{i} \mathrm{e}^{-\beta H_{o}})   \nonumber \\ &= \frac{2}{\beta}\int d \bm{x} f_{i}  \frac{\partial^{2}  H_{o}}{\partial x_{i} \partial t} \mathrm{e}^{-\beta H_{o}} . \label{eq:Secondterm}\end{align}
Here we have assumed that the boundary term vanishes at infinity. According to the above analysis, we can finally reduce the nonequilibrium constraint~(\ref{eq:Sabefore}) into the following simplified form:
\begin{align}  \mathcal{G}_{s} (\bm{f}) =&  \int d \bm{x} \left(     f_{i} \frac{\partial H_{o}}{\partial x_{i}} - \frac{1}{\beta} \frac{\partial f_{i}}{\partial x_{i}} \right)^{2}\mathrm{e}^{-\beta H_{o}}   \nonumber \\ &+ \frac{2}{\beta}\int d \bm{x} f_{i}  \frac{\partial^{2}  H_{o}}{\partial x_{i} \partial t} \mathrm{e}^{-\beta H_{o}} , \label{eq:Saafter}\end{align}
which is our first central result. Here the requirement about the free energy information has been eliminated. According to different restrictions on the auxiliary field, we first choose a proper trial function $\bm{f}(\bm{x},t)$. Then, substituting the trial function into the nonequilibrium constraint~(\ref{eq:Saafter}) and applying the variational procedure, we can find the best possible auxiliary field $\bm{f}^{a}(\bm{x},t)$ and thereby approximately realize shortcuts to isothermality. We dub such a variational scheme the ``variational shortcut to isothermality''.

For simple forms of $H_{o}(\bm{x},\lambda(t))$, calculating the integral in the constraint~(\ref{eq:Saafter}) is very straightforward. However, for complex systems where the integral in the nonequilibrium constraint~(\ref{eq:Saafter}) can not be accurately calculated, we may refer to some techniques for approximating the integral, such as the saddle-point approximation~\cite{Butler2007}. In order to get better approximation, we make further transformations of the nonequilibrium constraint~(\ref{eq:Saafter}). By using integration by parts, we can derive that (see Appendix~\ref{SSec-three} for details)
\begin{align}  \mathcal{G}_{s} (\bm{f})   = \int d \bm{x} W \mathrm{e}^{-\beta H_{o}} , \label{eq:Saafterfinal}\end{align}
with
\begin{align}    W(\bm{x},t)= \frac{1}{\beta^{2}}  \frac{\partial f_{i}}{\partial x_{j} } \frac{\partial f_{j}}{\partial x_{i} }  +  \frac{1}{\beta} f_{i} f_{j} \frac{\partial^{2} H_{o}}{ \partial x_{i} \partial x_{j} } + \frac{2}{\beta} f_{i}  \frac{\partial^{2}  H_{o}}{\partial x_{i} \partial t} . \label{eq:wfunction}\end{align}
Then, applying the saddle-point approximation to the integral~(\ref{eq:Saafterfinal}), we can obtain (see Appendix~\ref{SSec-three})
\begin{align} \mathcal{G}_{s} (\bm{f})  \approx \sum_{ m  } W(\bm{x}^{m},t)e^{-\beta H_{o}(\bm{x}^{m},\lambda)}\prod_{i=1}^{N}\sqrt{\frac{2\pi}{\beta \Lambda_{i}(\bm{x}^{m},t)}}, \label{eq:msaddleapprox1}\end{align}
where $\Lambda_{i}$ is an eigenvalue of the Hessian matrix $\bm{D}$ with $D_{jk} \equiv (\partial^{2} H_{o}/\partial x_{j} \partial x_{k})|_{\bm{x}=\bm{x}^{m}}$. Eq.~(\ref{eq:msaddleapprox1}) is our second central result. Here $ \bm{x}^{m} $ represents one of the minimum points of the function $H_{o}(\bm{x},\lambda)$. When applying the saddle-point approximation, we have assumed that the integral function $W \mathrm{e}^{-\beta H_{o}}$ is largely peaked around the points $ \bm{x}^{m} $. Please see Appendix~\ref{SSec-three} for detailed explanation about this assumption.

\section{Application\label{nonwore}}
Considering an underdamped Brownian particle controlled by an original potential $U_{o}(\bm{q},\lambda(t))$ and a momentum-dependent auxiliary potential $ U_{a}(\bm{q},\bm{p},t)$ with $\bm{q}=(q_{1},q_{2},\cdots,q_{N})$ and $\bm{p}=(p_{1},p_{2},\cdots,p_{N})$ denoting coordinate and momentum of the particle, respectively. The motion of the particle is governed by the modified Langevin equation~\cite{Geng2017}
\begin{align}   &\dot{q}_{i} = \frac{p_{i}}{m} +  \frac{\partial U_{a}}{\partial p_{i}},  \nonumber \\ &\dot{p}_{i}= - \frac{\partial U_{o}}{\partial q_{i}}-\frac{\partial U_{a}}{\partial q_{i}}-\gamma \left(\frac{p_{i}}{m} +   \frac{\partial U_{a}}{\partial p_{i}}  \right)+\xi_{i}(t),\label{eq:underLange}\end{align}
with $m$ being the mass of the particle. 
$\gamma$ represents the coefficient of friction and $\bm{\xi}=(\xi_{1},\xi_{2},\cdots,\xi_{N})$ denotes the standard Gaussian white noise satisfying $\langle \xi_{i}(t) \rangle=0$ and $\langle \xi_{i}(t) \xi_{j}(t^\prime) \rangle=2\gamma k_{B} T \delta_{ij} \delta(t-t^\prime)$. We propose the following trial form for the auxiliary potential
\begin{equation} U_{a}(\bm{q},\bm{p},t)=\dot{\lambda}(t)\left \{ [ s(\lambda(t))q_{i} +  u_{i}(\lambda(t)) ] p_{i}+ v(\bm{q},\lambda(t))         \right\},    \label{eq:generalU1}\end{equation}
where $s(\lambda(t))$, $ \bm{u}(\lambda(t))$, and $v(\bm{q},\lambda(t))$ are undetermined functions. To ensure that $U_{a}$ vanishes at the beginning and end of the driving process, we impose the boundary conditions $\dot{\lambda}(0)=\dot{\lambda}(\tau)=0$. The cross term $q_{i} p_{i}$ in the auxiliary potential~(\ref{eq:generalU1}), which is very hard to be realized in experiment~\cite{Guery2019,Geng2017}, can be eliminated by introducing a change of variables (see Appendix~\ref{SSec-four} for details).

As an illustrative example, we consider a one-dimensional double-well potential
\begin{equation} U_{o}(q,\lambda(t))=kq^{4}-\lambda(t)q^{2},  \label{eq:Sunpotential}\end{equation}
where $k$ is a constant. The Brownian motion of a particle in the double-well potential~(\ref{eq:Sunpotential}) is widely used to describe noise-driven motion in a variety of bistable physical and chemical systems~\cite{Saito1976,Hanggi1990,Coffey2004,Sun2003}. The evolution of the Brownian particle trajectory $\bm{x}=\{ q ~~  p  \}^{T}$ is then described by the Langevin equation
\begin{align} & \dot{q}=\frac{p}{m} +  \frac{\partial U_{a}}{\partial p}, \nonumber \\ & \dot{p}= -\frac{\partial U_{o}}{\partial q}  -  \frac{\partial U_{a}}{\partial q}  - \gamma \left( \frac{p}{m} +  \frac{\partial U_{a}}{\partial p}  \right)+ \xi(t). \label{eq:underLangevinone}\end{align}
Comparing it with the general motion equation~(\ref{eq:modifymoeq}), we obtain the corresponding relations
\begin{equation} \bm{f}^{o}(q,p,t)= \left \{ \frac{p}{m}  ~~  -\frac{\partial U_{o}}{\partial q}    -  \frac{\gamma}{m}p + \xi(t)   \right \}^{T},  \label{eq:underdamG0}\end{equation}
and
\begin{equation} \bm{f}^{a}(q,p,t)= \left \{ \frac{\partial U_{a}}{\partial p}  ~~  -  \frac{\partial U_{a}}{\partial q}  - \gamma   \frac{\partial U_{a}}{\partial p}     \right \}^{T}.    \label{eq:underdamG1}\end{equation}
In the one-dimensional underdamped Brownian particle system, the nonequilibrium constraint~(\ref{eq:msaddleapprox1}) can be simplified to the form
\begin{align}   \mathcal{G}_{s} \approx \sum_{m}  W'(q_{m},t)\mathrm{e}^{-\beta U_{o}(q_{m},\lambda)} \sqrt{\frac{2\pi}{\beta \Lambda(q_{m},\lambda)}}    ,\label{eq:underdamaction2}\end{align}
with
\begin{align}   W'(q,t) = & \frac{1}{\beta^{2}}\left \langle \left( \frac{\partial f_{q}}{\partial q}  \right)^{2} \right \rangle_{p}  + \frac{1}{\beta^{2}}\left \langle \left( \frac{\partial f_{p}}{\partial p}  \right)^{2} \right \rangle_{p}+ \frac{2}{\beta^{2}}\left \langle  \frac{\partial f_{p}}{\partial q}  \frac{\partial f_{q}}{\partial p}\right \rangle_{p} +  \frac{1}{\beta}\left \langle  \frac{\partial^{2} U_{o}}{\partial q^{2}} f_{q}^{2}   \right \rangle_{p}  \nonumber \\ & +  \frac{1}{\beta}\left \langle f_{p}^{2}   \right \rangle_{p} +\frac{2}{\beta}\left \langle f_{q}\frac{\partial^{2} U_{o}}{\partial q \partial t} \right \rangle_{p} .\label{eq:underdamactionw}\end{align}
Here $\left \langle  \cdots  \right \rangle_{p} \equiv  \int_{-\infty}^{+\infty} \cdots e^{-\beta p^{2}/2} dp$ represents the integral in the momentum space. It can be calculated directly without using the saddle-point approximation. $\bm{f} \equiv \{ f_{q}  ~~ f_{p} \}^{T}$ denotes an approximation to $\bm{f}^{a} \equiv \{ f_{q}^{a} ~~ f_{p}^{a} \}^{T}$.
According to the trial form~(\ref{eq:generalU1}), we assume that the auxiliary potential takes the form
\begin{align}  U_{a}(q,p,t)= \dot{\lambda}(t)[b_{6}^{*}(t)qp+b_{5}^{*}(t)p +b_{4}^{*}(t)q^{4}+ b_{3}^{*}(t)q^{3} + b_{2}^{*}(t)q^{2}+b_{1}^{*}(t)q ],              \label{eq:underdampreU1o}\end{align}
where $b_{1}^{*}(t)$, $b_{2}^{*}(t)$, $b_{3}^{*}(t)$, $b_{4}^{*}(t)$, $b_{5}^{*}(t)$, and $b_{6}^{*}(t)$ are undetermined parameters. Therefore, the approximate auxiliary field should take the form
\begin{align} \bm{f}= \left \{ \dot{\lambda}(b_{6}q+b_{5})  ~~  -\dot{\lambda}[b_{6}(p+\gamma q)+ \gamma b_{5}+4b_{4}q^{3}  +3b_{3}q^{2}+2b_{2}q+b_{1}]   \right \}^{T},    \label{eq:underdamA}\end{align}
with $b_{1}(t)$, $b_{2}(t)$, $b_{3}(t)$, $b_{4}(t)$, $b_{5}(t)$, and $b_{6}(t)$ being approximations to the corresponding parameters $b_{1}^{*}(t)$, $b_{2}^{*}(t)$, $b_{3}^{*}(t)$, $b_{4}^{*}(t)$, $b_{5}^{*}(t)$, and $b_{6}^{*}(t)$. Substituting the form~(\ref{eq:underdamA}) into the nonequilibrium constraint~(\ref{eq:underdamaction2}) and then minimizing it over the parameters, we can derive that the best possible auxiliary potential follows
\begin{equation} U_{a}(q,p,t)= \dot{\lambda}\left(  \frac{\beta \lambda}{2\beta\lambda^{2}+3k}qp+ b_{4}^{*}q^{4} + b_{2}^{*}q^{2} \right),          \label{eq:underdambestpoten}\end{equation}
where the undetermined parameters $b_{2}^{*}$ and $b_{4}^{*}$ should satisfy the relation
\begin{equation} b_{2}^{*}+\frac{\lambda}{k} b_{4}^{*} =-\frac{\gamma\beta\lambda}{4\beta\lambda^{2}+6k}.         \label{eq:pararelation}\end{equation}
Equation~(\ref{eq:pararelation}) gives us flexibility for choosing the parameters $b_{2}^{*}$ and $b_{4}^{*}$. We can compare the form of the auxiliary potential~(\ref{eq:underdambestpoten}) with the one in the overdamped situation and then derive that $b_{2}^{*}=-3\gamma\beta \lambda/(8\beta\lambda^{2}+12k)$ and $b_{4}^{*}=\gamma\beta k/(8\beta\lambda^{2}+12k)$. See Appendix~\ref{SSec-five} for detailed discussion. Therefore, the best possible auxiliary potential takes the form
\begin{equation} U_{a}(q,p,t)=\frac{\beta \dot{\lambda}}{8\beta \lambda^{2}+12k} \left( 4\lambda qp+\gamma kq^{4}-3\gamma \lambda q^{2}   \right).    \label{eq:doubleU1}\end{equation}

\begin{figure}[!htp]
\centering
\includegraphics[width=.8\linewidth]{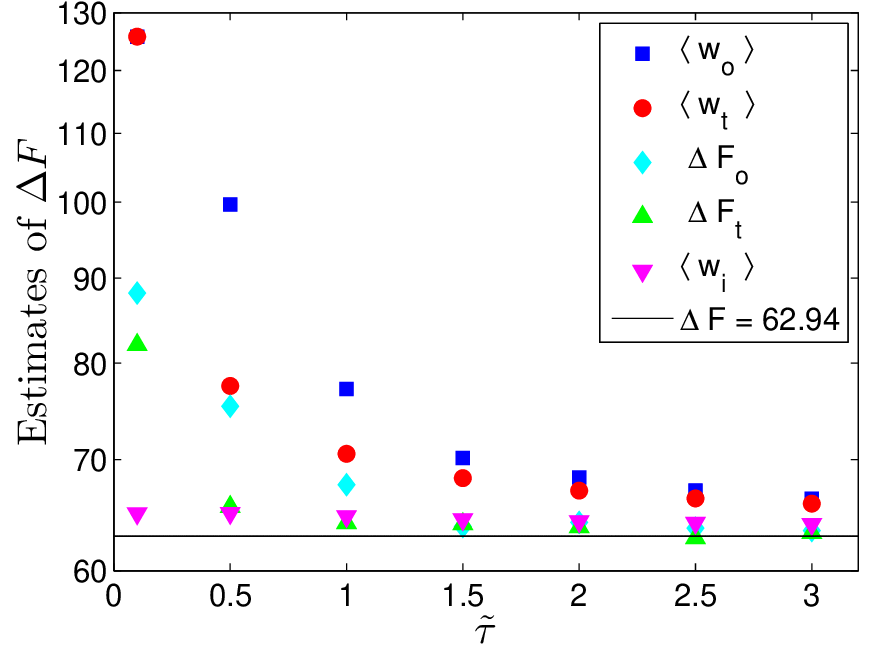}
 \caption{Comparison of estimates of $\Delta F$ for $\alpha=1.0$. $\langle w_{o} \rangle$ (squares) and $\Delta F_{o}$ (diamonds) represent the estimates from the mean work and the Jarzynski equality in the process driven by $U_{o}$ only. $\langle w_{t} \rangle$ (circles), $\Delta F_{t}$ (upper triangles), and $\langle w_{i} \rangle$ (lower triangles) represent the estimates from the mean work, the Jarzynski equality, and the intrinsic work~(\ref{eq:meaniw}) in the process driven by $U_{o}$ and $U_{a}$. The solid line represents the theoretical value~\cite{Sun2003}, $\Delta F=62.94$. The estimates of $\Delta F$ are shown on a logarithmic scale.}
 \label{fig1}
\end{figure}

We simulate the motion of an underdamped Brownian particle in the potential~(\ref{eq:Sunpotential}) and add the auxiliary potential~(\ref{eq:doubleU1}) to approximately realize shortcuts to isothermality. The dimensionless driving protocol is chosen to be $\tilde{\lambda}(t)=8[1+\cos (\pi t/\tau)]$ with $\tilde{\lambda} \equiv \lambda /\sqrt{k k_{B}T}$. The influence of the particle inertia is determined by a parameter $\alpha \equiv \tau_{p} / \tau_{q}$ with $\tau_{p} \equiv m/\gamma$ and $\tau_{q} \equiv \gamma / \sqrt{k k_{B}T}$ denoting two characteristic times of the system. The simulations are performed for dimensionless driving times $\tilde{\tau} \equiv \tau/ \tau_{p}$ ranging from $0.1$ to $3.0$. Details of the simulation are attached in Appendix~\ref{SSec-seven}. We use Eq.~(\ref{eq:meaniw}) to estimate the free energy difference $\Delta F$. As shown in Fig.~\ref{fig1}, the results are compared with the estimates given by the mean work and the Jarzynski equality. Here, $\langle w_{o} \rangle \equiv \int_{0}^{\tau} \left \langle\frac{\partial U_{o}(\bm{x},t)}{\partial t}\right \rangle dt$ and $ \Delta F_{o} \equiv -\beta^{-1}\ln \left\langle \mathrm{ e}^{-\beta w_{o}}  \right\rangle$ respectively represent the estimates from the mean work and the Jarzynski equality in the process driven by $U_{o}$ only, while $\langle  w_{t} \rangle \equiv \int_{0}^{\tau}\left( \left\langle \frac{\partial U_{o} }{\partial t} \right\rangle+\left\langle \frac{\partial U_{a} }{\partial t}\right\rangle \right)dt$ and $\Delta F_{t} \equiv -\beta^{-1}\ln \left \langle  \mathrm{e}^{-\beta w_{t}} \right \rangle$ respectively represent the estimates from the mean work and the Jarzynski equality in the process driven by $U_{o}$ and $U_{a}$.
\begin{figure}[!htp]
\centering
\includegraphics[width=.8\linewidth]{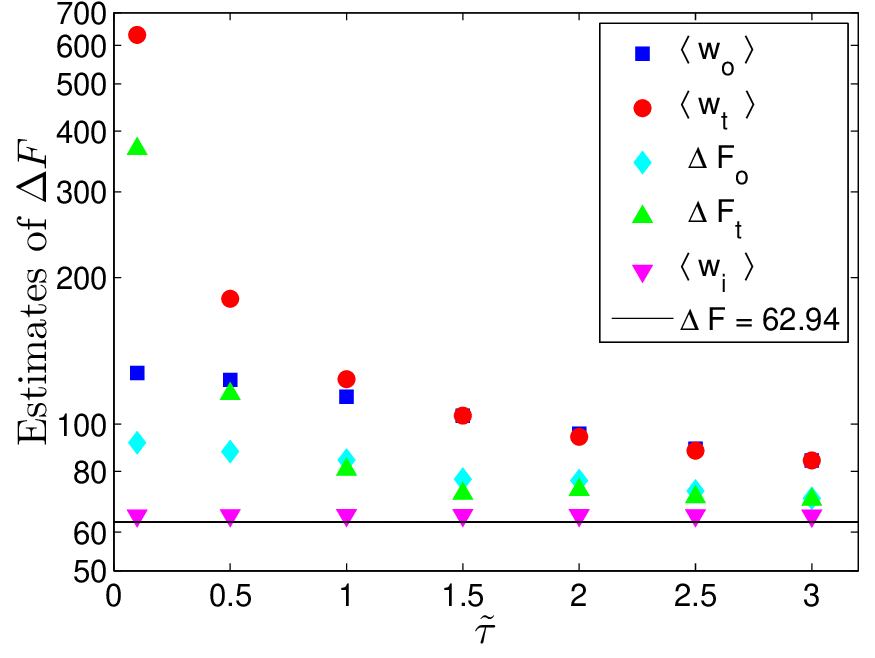}
 \caption{Comparison of estimates of $\Delta F$ for $\alpha=0.1$. The caption for Fig.~\ref{fig1} applies here.}
\label{fig2}
\end{figure}

Figure~\ref{fig1} shows that Eq.~(\ref{eq:meaniw}) provides a remarkably accurate and stable estimates of $\Delta F$. Especially in short driving times, where dissipation is expected to be high, the estimates given by Eq.~(\ref{eq:meaniw}) largely outperforms the estimates given by the mean work and the Jarzynski equality. We also compares the estimates when the inertia is small ($\alpha =0.1$). As shown in Fig.~\ref{fig2}, the estimates given by Eq.~(\ref{eq:meaniw}) are superior to other estimates over the entire range of driving times. These observations show that the variational shortcut to isothermality is promising to provide a reliable scheme for high-efficiency free energy estimation. This is our third central result.
\begin{figure}[!htp]
\centering
\includegraphics[width=.8\linewidth]{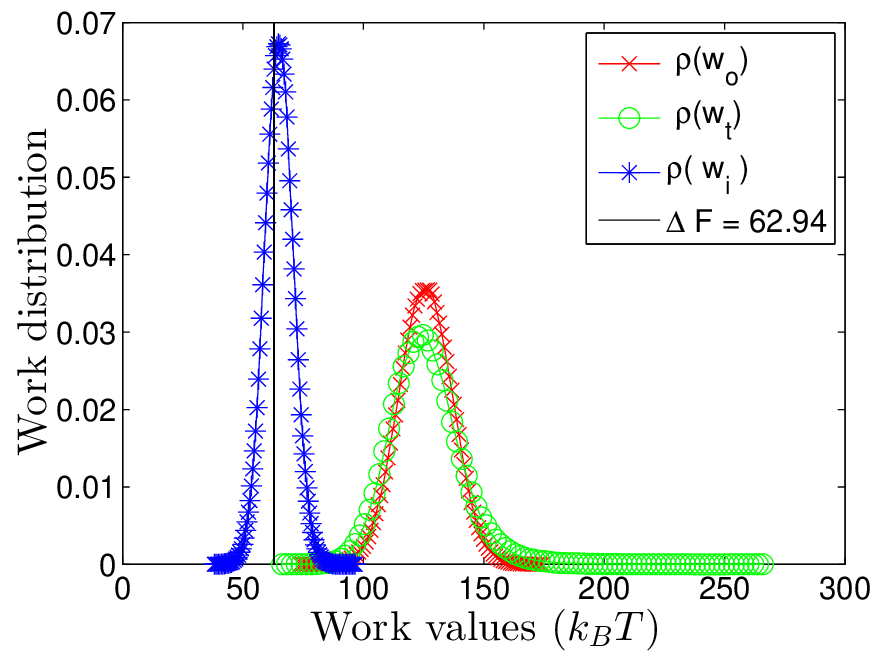}
 \caption{Comparison of different work distributions for $\alpha=1.0$. The driving time is $0.1$. $\rho(w_{o})$ (crosses) denotes the distribution of the total trajectory work $w_{o}$ in the process driven by $U_{o}$ only. $\rho(w_{t})$ (open circles) and $\rho(w_{i})$ (asterisks) denote the distributions of the total trajectory work $w_{t}$ and the intrinsic trajectory work $w_{i}$ in the process driven by $U_{o}$ and $U_{a}$. }
 \label{fig3}
\end{figure}

Figure~\ref{fig3} shows the comparison of different trajectory-work distributions for $\alpha=1.0$. Here we choose a short driving time, $\tilde{\tau}=0.1$. As shown in Fig.~\ref{fig3}, the distribution of the intrinsic trajectory work $\rho(w_{i})$ is sharply centered around the theoretical value of $\Delta F$ while the peaks of the total trajectory-work distributions $\rho(w_{o})$ and $\rho(w_{t})$ deviate far from the theoretical value of $\Delta F$. Besides, $\rho(w_{o})$ and $\rho(w_{t})$ take much broader forms than $\rho(w_{i})$. These observations imply that compared with the mean work and the Jarzynski equality, the intrinsic work equality~(\ref{eq:meaniw}) may allow us to obtain a reliable estimate of the free energy difference with a small number of trajectories. This is a superiority for the intrinsic work equality~(\ref{eq:meaniw}) when we are dealing with practical systems in which only a small number of samples are available.

\section{Conclusion and Discussion\label{disclu}}
Enlightened by the idea of the Gauss principle of least constraint, we have developed the variational shortcut to isothermality, which can approximately accelerate the conventional isothermal process for complex systems. A key advantage of this variational method is that it allows us to obtain the best possible auxiliary potential for shortcuts to isothermality without resort to the free energy information. Combined with the linear nonequilibrium equality~(\ref{eq:meaniw}), the variational method can be used to estimate the free energy difference. We have applied our method to an underdamped Brownian particle moving in a double-well potential. The simulations show that our method can accurately estimate the free energy difference with high efficiency. A potential future direction is to test our method on more complex multidimensional systems.

Considering the experimental feasibility, we have proposed a trial form~(\ref{eq:generalU1}) for the auxiliary potential of the underdamped Brownian particle system. In numerical simulations, we can assume a trial form with high-order couplings between the coordinate and the momentum. The variational shortcut to isothermality is still applicable in this situation and may provide a more accurate estimate of the free energy difference.

Here we have stressed the application of our variational method in accelerating the isothermal process and estimating the free energy difference. Related problems are that of importance sampling~\cite{Mazonka1998,Allen2005,Kundu2011}, shortcuts to stochastic near-adiabatic pumping~\cite{Funo2020,Takahashi2020}, preprocessing strategies before heating and cooling~\cite{Lu2017,Gal2020,Kumar2020}, thermodynamic controls~\cite{Deffner2020}, and so on. Similar to shortcuts to isothermality, the dynamics of importance sampling is also modified so as to simulate rare events of the original dynamics more frequently~\cite{Kundu2011}. In spite of the different targets and ways of modifying the dynamics, the idea of our variational method is promising to be extended to these problems.

\emph{Acknowledgement.}--The authors acknowledge valuable discussions with H. J. Zhou, C. P. Sun, and H. Dong. G. L. is supported by the National Natural Science Foundation of China under Grants No. 12047549 and No. 11947302. Z. C. T. is supported by the National Natural Science Foundation of China under Grants No. 11975050 and No. 11675017.

\appendix

\section{Derivation of the evolution equation for the auxiliary field, Eq.~(\ref{eq:G1evoequ})\label{SSec-one}}
Consider a system following the motion equation
\begin{equation} \dot{x}_{i} = f_{i}^{o}(\bm{x},t). \label{seq:moeq}\end{equation}
The evolution equation of the system distribution $\rho(\bm{x},t)$ can be formally written as
\begin{eqnarray}  \frac{\partial \rho}{\partial t} =  -\frac{\partial}{\partial x_{i}} (\dot{x}_{i} \rho) =  -\frac{\partial}{\partial x_{i}} (f_{i}^{o} \rho) . \label{seq:evoleq}\end{eqnarray}
If we consider $\bm{ f}^{o}(\bm{x},t)$ containing both deterministic and stochastic parts (such as the Langevin dynamics), the time evolution equation~(\ref{seq:evoleq}) will be different for each realization of the stochastic parts~\cite{Reichl2009}. After averaging over the stochastic parts, we can formally derive the evolution equation of the observable probability:
\begin{eqnarray}  \frac{\partial \rho}{\partial t} = \hat{L}_{o} \rho , \label{seq:observaevoleq}\end{eqnarray}
where $\hat{L}_{o} \equiv \hat{L}_{o}(\bm{x},t)$ represents the evolution operator.
If we add an auxiliary potential $U_{a}(\bm{x},t)$ to the original Hamiltonian, the motion equation is modified to the form
\begin{equation} \dot{x}_{i} = f_{i}^{o}(\bm{x},t)+  f_{i}^{a}(\bm{x},t),\label{seq:moeqG1}\end{equation}
where the auxiliary field $\bm{ f}^{a}(\bm{x},t)$ depends on $U_{a}(\bm{x},t)$ and the dynamics we are considering. Since the ensemble average over the stochastic parts of $\bm{ f}^{o}(\bm{x},t)$ does not affect the deterministic field $\bm{ f}^{a}(\bm{x},t)$, we can formally derive the modified evolution equation as
\begin{eqnarray}  \frac{\partial \rho}{\partial t} = \hat{L}_{o} \rho -\frac{\partial}{\partial x_{i}} (f_{i}^{a}\rho ), \label{seq:observG1}\end{eqnarray}
which is just Eq.~(\ref{eq:observG1}) in the main text. When we adopt the strategy of shortcuts to isothermality, the system distribution will always stay in the instantaneous canonical distribution of $H_{o}(\bm{x},\lambda(t))$:
\begin{equation} \rho^{\mathrm{ieq}}(\bm{x},\lambda(t))= \mathrm{e}^{\beta[F(\lambda(t))-H_{o}(\bm{x},\lambda(t))]}. \label{seq:ieqdistri}\end{equation}
Substituting the instantaneous canonical distribution~(\ref{seq:ieqdistri}) into the modified evolution equation~(\ref{seq:observG1}), we can derive
\begin{equation} f_{i}^{a} \frac{\partial H_{o}}{\partial x_{i}} - \frac{1}{\beta} \frac{\partial f_{i}^{a}}{\partial x_{i}} =  \frac{d F}{d t} - \frac{\partial H_{o}}{\partial t}, \label{seq:G1evoequ}\end{equation}
which corresponds to Eq.~(\ref{eq:G1evoequ}) in the main text.

Since both $F(\lambda(t))$ and $H_{o}(\bm{x},\lambda(t))$ depend explicitly on time through the controlling parameter $\lambda(t)$, we can further derive
\begin{equation} f_{i}^{a} \frac{\partial H_{o}}{\partial x_{i}} - \frac{1}{\beta} \frac{\partial f_{i}^{a}}{\partial x_{i}} = \left ( \frac{d F}{d \lambda} - \frac{\partial H_{o}}{\partial \lambda} \right)\dot{\lambda}. \label{seq:G1evoequfur}\end{equation}
Comparing two sides of Eq.~(\ref{seq:G1evoequfur}), we find that $\bm{f}^{a}$ can be preassumed to take the form
\begin{equation}\bm{f}^{a}(\bm{x},t)= \dot{\lambda}(t)\bm{\nu}(\bm{x},\lambda(t)),\label{U1form}\end{equation}
with $\bm{\nu}(\bm{x},\lambda(t))$ being an undetermined function.

\section{Equivalence between the evolution equation~(\ref{eq:G1evoequ}) and the variational equation~(\ref{eq:ELequ})\label{SSec-two}}
We start from the definition of the function
\begin{equation} \mathcal{D}(\bm{f}) \equiv f_{i} \frac{\partial H_{o}}{\partial x_{i}} - \frac{1}{\beta} \frac{\partial f_{i}}{\partial x_{i}} +  \frac{\partial H_{o}}{\partial t}  -  \frac{d F}{d t},    \label{seq:functionalG1}\end{equation}
where $\bm{f}\equiv \bm{f}(\bm{x},t)$ represents an approximation to the exact auxiliary field $\bm{f}^{a}(\bm{x},t)$. If $\bm{f}=\bm{f}^{a}$, then $\mathcal{D}(\bm{f})=0$. For any forms of $\bm{f}$, we can derive that
\begin{align}   \int d \bm{x}  \mathcal{D}(\bm{f}) \mathrm{e}^{-\beta H_{o}}   =   -\frac{1}{\beta} \int d \bm{x}  \frac{\partial}{\partial x_{i}} (f_{i} \mathrm{e}^{-\beta H_{o}})      + \int d \bm{x} \left(   \frac{\partial H_{o}}{\partial t}  -  \frac{d F}{d t} \right)\mathrm{e}^{-\beta H_{o}} =0. \label{seq:averageM}\end{align}
Referring to the Gauss principle of least constraint~\cite{Gauss1829}, we define a functional
\begin{equation}  \mathcal{G} (\bm{f}) \equiv  \int d\bm{x} \mathcal{D}^{2}(\bm{f}) \mathrm{e}^{-\beta H_{o}}  , \label{seq:Sactionn}\end{equation}
as a nonequilibrium ``constraint'' on the auxiliary field $\bm{f}$. If the form of the auxiliary field is free from restrictions, the nonequilibrium constraint~(\ref{seq:Sactionn}) will be minimized whenever $\bm{f}$ satisfies:
\begin{equation} \frac{\delta \mathcal{G}}{\delta \bm{f}}=0 \Rightarrow\left. \nabla \mathcal{D}\right|_{\bm{f}=\bm{f}^{a}} =0 , \label{seq:eularlangevin}\end{equation}
which then implies
\begin{equation} \left. \mathcal{D}\right|_{\bm{f}=\bm{f}^{a}}=C(t) \label{seq:Mequation}\end{equation}
with $C(t)$ being a time-dependent parameter. Because of the property~(\ref{seq:averageM}), we can derive $C(t)=0$, i.e.,
\begin{equation} \left.\mathcal{D}\right|_{\bm{f}=\bm{f}^{a}}= f_{i}^{a} \frac{\partial H_{o}}{\partial x_{i}} - \frac{1}{\beta} \frac{\partial f_{i}^{a}}{\partial x_{i}} +  \frac{\partial H_{o}}{\partial t}  -  \frac{d F}{d t}=0 ,\label{seq:Mequation0}\end{equation}
which, as anticipated, is just Eq.~(\ref{eq:G1evoequ}) in the main text. Therefore, unrestricted minimization of the nonequilibrium constraint~(\ref{seq:Sactionn}) is mathematically equivalent to solving Eq.~(\ref{eq:G1evoequ}). If restrictions prevent the free choice of $\bm{f}$, we can still minimize the nonequilibrium constraint~(\ref{seq:Sactionn}) under the given restrictions.

\section{Applying the saddle-point approximation to the nonequilibrium constraint\label{SSec-three}}
Consider an integral of the form
\begin{align}  \int_{y_{0}}^{y_{1}} dy w(y)e^{Ag(y)}, \label{seq:normalintegral}\end{align}
where $w(y)$ and $g(y)$ are some real functions and $A>0$ is a parameter. For large values of $A$, the integral~(\ref{seq:normalintegral}) is completely dominated by the peaks with each peak located at a maximum of $g(y)$. Without loss of generality, let us first assume that $y_{m}$ is the only maximum point of $g$ in the interval $(y_{0},y_{1})$. Changing the integral variable according to $y=y_{m}+z/\sqrt{A}$ and then expanding $Ag(y)$ in power of $z$, we have
\begin{align}  Ag(y) = Ag(y_{m}) + \frac{ z^{2}}{2} \left. \frac{d^{2} g(y)}{dy^{2}}\right|_{y=y_{m}}  + O\left( \frac{1}{\sqrt{A}}  \right). \label{seq:agzpower}\end{align}
Here the first-derivative term is missing because $y_{m}$ is the maximum of $g$. In the exponential form, we can further derive that
\begin{align}  \mathrm{exp}(Ag(y)) =  \mathrm{exp}\left(Ag(y_{m})+\frac{ z^{2}}{2} \left. \frac{d^{2} g(y)}{dy^{2}}\right|_{y=y_{m}}\right)   \left(1 + O\left( \frac{1}{\sqrt{A}}  \right)  \right). \label{seq:exponagzpower}\end{align}
Assuming that $w(y_{m}) \ne  0$, we can similarly expand $w(y)$ in power of $z$:
\begin{align}  w(y)=w(y_{m})\left( 1+ O\left( \frac{1}{\sqrt{A}}  \right) \right). \label{seq:wzpower}\end{align}
Substituting Eqs.~(\ref{seq:exponagzpower}) and~(\ref{seq:wzpower}) into~(\ref{seq:normalintegral}), we have
\begin{align}  \int_{y_{0}}^{y_{1}} dy w(y)e^{Ag(y)} = \frac{w(y_{m})e^{Ag(y_{m})}}{\sqrt{A}}\int_{z_{0}}^{z_{1}} dz \left[\mathrm{exp} \left( \frac{z^{2}}{2} \left. \frac{d^{2} g(y)}{dy^{2}}\right|_{y=y_{m}} \right)  \left( 1+ O\left( \frac{1}{\sqrt{A}}  \right) \right) \right]. \label{seq:integralpower}\end{align}
In the large $A$ limit, $z_{0}$ and $z_{1}$ will tend to $-\infty$ and $+\infty$, respectively. Therefore, we can give the saddle-point approximation:
\begin{align}  \int_{y_{0}}^{y_{1}} dy w(y)e^{Ag(y)} \approx w(y_{m})e^{Ag(y_{m})}\sqrt{\frac{2\pi}{-A\left. d^{2} g(y)/dy^{2}\right|_{y=y_{m}}}} . \label{seq:saddleapprox1}\end{align}
If there are multiple maxima of $g(y)$ in the integral interval $(y_{0},y_{1})$, we can divide the interval into smaller intervals according to the location of each maximum point. The integral~(\ref{seq:normalintegral}) equals to the sum of the approximation in each small interval.

In the following, we apply the saddle-point approximation to the integral in the nonequilibrium constraint. Starting from the nonequilibrium constraint~(\ref{eq:Saafter}), we can derive that
\begin{align}  \mathcal{G}_{s} (\bm{f}) = & \int d \bm{x} f_{i}f_{j} \frac{\partial H_{o}}{\partial x_{i}} \frac{\partial H_{o}}{\partial x_{j}}\mathrm{e}^{-\beta H_{o}}   + \frac{1}{\beta^{2}}\int d \bm{x}  \frac{\partial f_{i}}{\partial x_{i}} \frac{\partial f_{j}}{\partial x_{j}}\mathrm{e}^{-\beta H_{o}} \nonumber \\ &-  \frac{2}{\beta}\int d \bm{x} f_{i}\frac{\partial H_{o}}{\partial x_{i}} \frac{\partial f_{j}}{\partial x_{j}} \mathrm{e}^{-\beta H_{o}} + \frac{2}{\beta}\int d \bm{x} f_{i}  \frac{\partial^{2}  H_{o}}{\partial x_{i} \partial t} \mathrm{e}^{-\beta H_{o}}  . \label{seq:Saafter11}\end{align}
If we apply the saddle-point approximation directly to the integral in the nonequilibrium constraint~(\ref{seq:Saafter11}), the first term and the third term will vanish since there are first-order derivative of $ H_{o} $ in them. In order to get a better approximation, we make further transformations to the nonequilibrium constraint~(\ref{seq:Saafter11}). Applying integration by parts to the first term, we can derive that
\begin{align}  & \int d \bm{x} f_{i}f_{j} \frac{\partial H_{o}}{\partial x_{i}} \frac{\partial H_{o}}{\partial x_{j}}\mathrm{e}^{-\beta H_{o}} \nonumber \\ =&-\frac{1}{\beta} \int d \bm{x}  f_{i}f_{j} \frac{\partial H_{o}}{\partial x_{i}} \frac{\partial (\mathrm{e}^{-\beta H_{o}})}{\partial x_{j}} \nonumber \\ =& \frac{1}{\beta} \int d \bm{x} \left[ \frac{\partial }{\partial x_{j}}  \left(f_{i}f_{j} \frac{\partial H_{o}}{\partial x_{i}} \right)  \right] \mathrm{e}^{-\beta H_{o}} \nonumber \\ =& \frac{1}{\beta} \int d \bm{x} \left(  f_{j} \frac{\partial f_{i}}{\partial x_{j}} \frac{\partial H_{o}}{\partial x_{i}} +  f_{i}f_{j} \frac{\partial^{2} H_{o}}{\partial x_{i} \partial x_{j}} + f_{i} \frac{\partial H_{o}}{\partial x_{i}}  \frac{\partial f_{j}}{\partial x_{j}}     \right) \mathrm{e}^{-\beta H_{o}} . \label{seq:Saafterfirst}\end{align}
Here we have also assumed that the boundary terms vanish at infinity. We can similarly obtain that
\begin{align}   \frac{1}{\beta} \int d \bm{x}   f_{j} \frac{\partial f_{i}}{\partial x_{j}} \frac{\partial H_{o}}{\partial x_{i}} \mathrm{e}^{-\beta H_{o}}  = \frac{1}{\beta^{2}}\int d \bm{x} \left(  \frac{\partial f_{i}}{\partial x_{j}} \frac{\partial f_{j}}{\partial x_{i}} + f_{j} \frac{\partial^{2} f_{i}}{\partial x_{i}\partial x_{j} }   \right)  \mathrm{e}^{-\beta H_{o}}, \label{seq:Saaftersecond}\end{align}
and
\begin{align}   \frac{1}{\beta} \int d \bm{x}   f_{i} \frac{\partial H_{o}}{\partial x_{i}}  \frac{\partial f_{j}}{\partial x_{j}}     \mathrm{e}^{-\beta H_{o}} = \frac{1}{\beta^{2}} \int d \bm{x} \left(  \frac{\partial f_{i}}{\partial x_{i}} \frac{\partial f_{j}}{\partial x_{j}}  + f_{j} \frac{\partial^{2} f_{i}}{\partial x_{i}\partial x_{j} }  \right ) \mathrm{e}^{-\beta H_{o}}. \label{seq:Saafterthird}\end{align}
Substituting Eqs.~(\ref{seq:Saafterfirst}),~(\ref{seq:Saaftersecond}), and~(\ref{seq:Saafterthird}) into~(\ref{seq:Saafter11}), we can finally derive that
\begin{align}  \mathcal{G}_{s} (\bm{f})  =&  \frac{1}{\beta^{2}} \int d \bm{x}  \frac{\partial f_{i}}{\partial x_{j} } \frac{\partial f_{j}}{\partial x_{i} }  \mathrm{e}^{-\beta H_{o}} +   \frac{1}{\beta} \int d \bm{x} f_{i} f_{j} \frac{\partial^{2} H_{o}}{ \partial x_{i} \partial x_{j} }   \mathrm{e}^{-\beta H_{o}}   + \frac{2}{\beta}\int d \bm{x} f_{i}  \frac{\partial^{2}  H_{o}}{\partial x_{i} \partial t} \mathrm{e}^{-\beta H_{o}} \nonumber \\ =& \int d \bm{x} W \mathrm{e}^{-\beta H_{o}} , \label{seq:Saafterfinal}\end{align}
with
\begin{align}    W(\bm{x},t)= \frac{1}{\beta^{2}}  \frac{\partial f_{i}}{\partial x_{j} } \frac{\partial f_{j}}{\partial x_{i} }  +  \frac{1}{\beta} f_{i} f_{j} \frac{\partial^{2} H_{o}}{ \partial x_{i} \partial x_{j} } + \frac{2}{\beta} f_{i}  \frac{\partial^{2}  H_{o}}{\partial x_{i} \partial t} . \label{seq:wfunction}\end{align}
Here we have cancelled out the terms containing the first-order derivative of $H_{o}$ in the nonequilibrium constraint.

Without loss of generality, we assume that the Hamiltonian function $ H_{o}$ has only one minimum located at $\bm{x}^{m}$. We use $\Delta \equiv E_{min}-E_{max} $ to denote the difference between the minimum $E_{min}$ of the function $ H_{o}$ and its adjoining maximum $E_{max}$. Then the exponential term $-\beta H_{o}$ can be transformed into
\begin{align}  -\beta H_{o} = A \tilde{H}_{o},  \label{seq:exponentterm}\end{align}
where $A\equiv -\beta \Delta$ and $\tilde{H}_{o} \equiv H_{o}/\Delta$. When $A \gg 1$, the saddle-point approximation can be applied to the integral~(\ref{seq:Saafterfinal}). Changing the integral variable according to $\bm{x}=\bm{x}^{m}+\bm{z}/\sqrt{A}$, we can expand $A  \tilde{H}_{o}$ in power of $\bm{z}$:
\begin{align}  A  \tilde{H}_{o}(\bm{x},\lambda) =  A\tilde{H}_{o}(\bm{x}^{m},\lambda) + \frac{ z_{i}z_{j}}{2} \left. \frac{\partial^{2} \tilde{H}_{o}}{\partial x_{i}\partial x_{j}}\right|_{\bm{x}=\bm{x}^{m}} + O\left( \frac{1}{\sqrt{A}}  \right). \label{seq:ah0power}\end{align}
Similarly, we can derive that
\begin{align}  \mathrm{exp}(A  \tilde{H}_{o}) =   \mathrm{exp}\left(A\tilde{H}_{o}(\bm{x}^{m},\lambda) + \frac{ z_{i}z_{j}}{2} \left. \frac{\partial^{2} \tilde{H}_{o}}{\partial x_{i}\partial x_{j}}\right|_{\bm{x}=\bm{x}^{m}}\right)   \left(1 + O\left( \frac{1}{\sqrt{A}} \right)  \right), \label{seq:mexponagzpower}\end{align}
and
\begin{eqnarray}  W(\bm{x},t)=W(\bm{x}^{m},t)\left( 1+  O\left( \frac{1}{\sqrt{A}}  \right) \right). \label{seq:mwzpower}\end{eqnarray}
Substituting Eqs.~(\ref{seq:mexponagzpower}) and~(\ref{seq:mwzpower}) into~(\ref{seq:Saafterfinal}), we have
\begin{align}   \mathcal{G}_{s} (\bm{f}) =& \frac{W(\bm{x}^{m},t)e^{A  \tilde{H}_{o}(\bm{x}^{m},\lambda)}}{\sqrt{A}} \int d\bm{z} \left[\mathrm{exp} \left( \frac{ z_{i}z_{j}}{2} \left. \frac{\partial^{2} \tilde{H}_{o}}{\partial x_{i}\partial x_{j}}\right|_{\bm{x}=\bm{x}^{m}} \right)  \left( 1+ O\left( \frac{1}{\sqrt{A}}  \right) \right) \right]. \label{seq:mintegralpower}\end{align}
In the large $A$ limit, we can give the saddle-point approximation:
\begin{align} \mathcal{G}_{s} (\bm{f})  \approx  W(\bm{x}^{m},t)e^{-\beta H_{o}(\bm{x}^{m},\lambda)}\prod_{i=1}^{n}\sqrt{\frac{2\pi}{\beta \Lambda_{i}(\bm{x}^{m},t)}}, \label{seq:msaddleapprox1}\end{align}
where $\Lambda_{i} $ is an eigenvalue of the Hessian matrix
\begin{align}
 \bm{D}= \left. \left( \begin{array}{cccc} \frac{\partial^{2} H_{o}}{\partial x_{1}^{2}} & \frac{\partial^{2} H_{o}}{\partial x_{1}\partial x_{2}} & \ldots &  \frac{\partial^{2} H_{o}}{\partial x_{1}\partial x_{n}} \\ \frac{\partial^{2} H_{o}}{\partial x_{2}\partial x_{1}} & \frac{\partial^{2} H_{o}}{\partial x_{2}^{2}} & \ldots & \frac{\partial^{2} H_{o}}{\partial x_{2}\partial x_{1}} \\  \vdots & \vdots & \ddots &  \vdots \\ \frac{\partial^{2} H_{o}}{\partial x_{n}\partial x_{1}} & \frac{\partial^{2} H_{o}}{\partial x_{n}\partial x_{2}} & \ldots & \frac{\partial^{2} H_{o}}{\partial x_{n}^{2}} \end{array}  \right) \right|_{\bm{x}=\bm{x}^{m}} .   \label{seq:hessianmatrix}
\end{align}
Similarly, if the Hamiltonian function $ H_{0}$ has multiple minima $\{ \bm{x}^{m} \}$ with $m=1,2,\cdots$, the integral~(\ref{seq:Saafterfinal}) will be the sum of the approximation around each minimum point:
\begin{align} \mathcal{G}_{s} (\bm{f})  \approx \sum_{m} W(\bm{x}^{m},t)e^{-\beta H_{o}(\bm{x}^{m},\lambda)}\prod_{i=1}^{n}\sqrt{\frac{2\pi}{\beta \Lambda_{i}(\bm{x}^{m},t)}}. \label{seq:msaddleapprox1sum}\end{align}
We have assumed that $A \gg 1$ when applying the saddle-point approximation~(\ref{seq:msaddleapprox1}). In a nutshell, this assumption means that the integral function $W \mathrm{e}^{-\beta H_{o}}$ is largely peaked around the minimum points $\bm{x}^{m} $.

\section{Fast-forward protocol in shortcuts to isothermality\label{SSec-four}}
We start from the modified Langevin equation
\begin{align}   & \dot{q}_{i} = \frac{p_{i}}{m} +  \frac{\partial U_{a}}{\partial p_{i}},  \nonumber \\ &\dot{p}_{i}= - \frac{\partial U_{o}}{\partial q_{i}}-\frac{\partial U_{a}}{\partial q_{i}}-\gamma \left(\frac{p_{i}}{m} +   \frac{\partial U_{a}}{\partial p_{i}}  \right)+\xi_{i}(t),\label{seq:underLange}\end{align}
where the auxiliary potential takes the trial form
\begin{equation} U_{a}(\bm{q},\bm{p},t)=\dot{\lambda}(t)\left \{ [ s(\lambda(t))q_{i} +  u_{i}(\lambda(t)) ] p_{i}+ v(\bm{q},\lambda(t))         \right\}.    \label{seq:generalU1}\end{equation}
Similar to the nonlocal term in shortcuts to adiabaticity, the cross term $q_{i}  p_{i}$ in the auxiliary potential~(\ref{seq:generalU1}) is hard to be realized in experiment~\cite{Guery2019,Geng2017}. We now introduce a change of variables that can effectively eliminate the cross term.

Substituting Eq.~(\ref{seq:generalU1}) into Eq.~(\ref{seq:underLange}), we can obtain
\begin{align}   \dot{q}_{i}=& \frac{p_{i}}{m} +  \dot{\lambda}(sq_{i}+u_{i}),  \nonumber \\ \dot{p}_{i}=& - \frac{\partial U_{o}}{\partial q_{i}}-\dot{\lambda}\left( sp_{i} + \frac{\partial v}{\partial q_{i}} \right)   -\gamma \left[\frac{p_{i}}{m} +   \dot{\lambda}(sq_{i}+u_{i})  \right]+\xi_{i}(t).\label{seq:underLangeU1}\end{align}
Consider the evolution of the observables
\begin{equation} Q_{i}=q_{i}, \quad P_{i}=p_{i}+m\dot{\lambda}(sq_{i}+u_{i})   , \label{seq:mapping}\end{equation}
along a trajectory governed by the Langevin equation~(\ref{seq:underLangeU1}). Taking time derivative of the observables, we obtain
\begin{align}   \dot{Q}_{i}=& \dot{q}_{i},  \nonumber \\ \dot{P}_{i}=& \dot{p}_{i}+m\ddot{\lambda}(sq_{i}+u_{i})+m\dot{\lambda}^{2}  \left( \frac{\partial s}{\partial \lambda}q_{i} +\frac{d u_{i}}{d \lambda} \right)  +\dot{\lambda}sp_{i}.\label{seq:mapping2}\end{align}
By applying the mapping relations~(\ref{seq:mapping}) and~(\ref{seq:mapping2}) into Eq.~(\ref{seq:underLangeU1}), we get
\begin{align}   & \dot{Q}_{i} = \frac{P_{i}}{m},   \nonumber \\ &\dot{P}_{i}= - \frac{\partial U_{o}}{\partial Q_{i}}+F_{i}^{a}-\gamma\frac{ P_{i}}{m} +\xi_{i}(t),\label{seq:modifyunla}\end{align}
with the auxiliary force
\begin{align}     F_{i}^{a}(\bm{Q},t)=-\dot{\lambda}\frac{\partial v}{\partial Q_{i}}+m\ddot{\lambda}(sQ_{i}+u_{i})+m\dot{\lambda}^{2}  \left( \frac{\partial s}{\partial \lambda}Q_{i}+\frac{d u_{i}}{d \lambda} \right). \label{seq:auxiliforce}\end{align}
Here $\ddot{\lambda}$ represents the second time derivative of $\lambda$.

Similar to the fast-forward protocol in shortcuts to adiabaticity~\cite{Masuda2010,Torrontegui2012,Martinez2016,Jarzynski2017,Patra2017}, Eq.~(\ref{seq:modifyunla}) can approximately realize a transition between two equilibrium states at the same temperature in finite time. Additional boundary conditions $\ddot{\lambda}(0)=\ddot{\lambda}(\tau)=0$ need to be satisfied by the driving protocol. In the intermediate driving process, the system will depart from the instantaneous equilibrium state. Since $\bm{F}^{a}$ is an explicit function of $\bm{Q}$ and $t$, it will generically be easier to implement in experiment than the momentum-dependent auxiliary potential~(\ref{seq:generalU1}).

%

\section{Applying the variational shortcut to isothermality to a Brownian particle in the overdamped situation\label{SSec-five}}

In the overdamped situation, the motion of the Brownian particle is governed by the Langevin equation
\begin{equation} \dot{q}= -\frac{1}{\gamma} \frac{\partial U_{o}}{\partial q} -\frac{1}{\gamma} \frac{\partial U_{a}}{\partial q} + \frac{1}{\gamma} \xi(t) . \label{seq:overLange}\end{equation}
Comparing Eq.~(\ref{seq:overLange}) with the general motion equation~(\ref{seq:moeqG1}), we can obtain the corresponding relations
\begin{equation} f^{o}(q,t)= -\frac{1}{\gamma} \frac{\partial U_{o}}{\partial q} + \frac{1}{\gamma} \xi(t),  \label{seq:overG0}\end{equation}
and
\begin{equation} f^{a}(q,t)= -\frac{1}{\gamma} \frac{\partial U_{a}}{\partial q}.    \label{seq:overG1}\end{equation}
Note that $f^{o}(q,t)$ contains both the deterministic part $-\gamma^{-1} \partial U_{o}/\partial q$ and the stochastic part $\gamma^{-1} \xi(t)$ while $f^{a}(q,t)$ is presupposed to be deterministic.

In the one-dimensional overdamped Brownian particle system, the nonequilibrium constraint~(\ref{eq:msaddleapprox1}) can be simplified to the form
\begin{align}   \mathcal{G}_{s} \approx \sum_{m}  W(q_{m},t)\mathrm{e}^{-\beta U_{o}(q_{m},\lambda)} \sqrt{\frac{2\pi}{\beta \Lambda(q_{m},\lambda)}}    ,\label{seq:overaction2}\end{align}
with
\begin{align}   W(q,t)=  \frac{1}{\beta^{2}} \left( \frac{\partial f}{\partial q} \right)^{2} +  \frac{1}{\beta} f^{2}\frac{\partial^{2} U_{o}}{\partial q^{2}} +\frac{2}{\beta} f  \frac{\partial^{2} U_{o}}{\partial q\partial t}     .\label{seq:overactionw}\end{align}
Here $\Lambda(q_{m},\lambda)=(\partial^{2} U_{0} / \partial q^{2})|_{q=q_{m}}$ with $q_{m}$ representing one of the minimum points of the function $U_{0}$.

Considering the double-well potential
\begin{equation} U_{o}(q,\lambda(t))=kq^{4}-\lambda(t)q^{2} , \label{seq:Sunpotential}\end{equation}
we can derive that there are two minimum points $q_{1}=\sqrt{\lambda/2k}$ and $q_{2}=-\sqrt{\lambda/2k}$. According to the form of the original potential~(\ref{seq:Sunpotential}), we assume that the auxiliary potential takes the form
\begin{equation}  U_{a}(q,t)=\dot{\lambda}(t)[a_{4}^{*}(t)q^{4}+ a_{3}^{*}(t)q^{3}+ a_{2}^{*}(t)q^{2}+a_{1}^{*}(t)q ],              \label{seq:overpreU1}\end{equation}
where $a_{1}^{*}(t)$, $a_{2}^{*}(t)$, $a_{3}^{*}(t)$, and $a_{4}^{*}(t)$ are undetermined parameters. Therefore, the approximate auxiliary field should take the form
\begin{equation}  f(x,t)=-\frac{\dot{\lambda}(t)}{\gamma}[4a_{4}(t)q^{3}+ 3a_{3}(t)q^{2}+ 2a_{2}(t)q+a_{1}(t) ],              \label{seq:overpreA}\end{equation}
where $a_{1}(t)$, $a_{2}(t)$, $a_{3}(t)$, and $a_{4}(t)$ are approximations to the corresponding parameters $a_{1}^{*}(t)$, $a_{2}^{*}(t)$, $a_{3}^{*}(t)$, and $a_{4}^{*}(t)$.
Substituting the trial form~(\ref{seq:overpreA}) into the nonequilibrium constraint~(\ref{seq:overaction2}) and then minimizing it over the parameters, we obtain
\begin{align}
    \bm{M}\left( \begin{array}{cccc}  a_{4}^{*} \\ a_{3}^{*} \\ a_{2}^{*} \\ a_{1}^{*}     \end{array}  \right) = \left( \begin{array}{cccc}  -\gamma \overline{q^{1}} \\ -\gamma  \overline{q^{2}} \\ -\gamma  \overline{q^{3}} \\ -\gamma  \overline{q^{4}}     \end{array}  \right), \label{seq:overarray2}
\end{align}
where
\begin{align}
 \bm{M} \equiv \left( \begin{array}{cccc} 24k\overline{q^{5}}-4\lambda \overline{q^{3}} & 18k\overline{q^{4}}-3\lambda\overline{q^{2}} & 12k\overline{q^{3}}-2\lambda\overline{q^{1}} &  6k\overline{q^{2}}-\lambda\overline{q^{0}} \\ 24 k\overline{q^{6}}-4 \lambda \overline{q^{4}}+\frac{6}{\beta}\overline{q^{2}} &  18 k\overline{q^{5}}-3 \lambda\overline{q^{3}}+\frac{3}{\beta}\overline{q^{1}} & 12 k\overline{q^{4}}-2 \lambda\overline{q^{2}}+\frac{1}{\beta}\overline{q^{0}} & 6 k\overline{q^{3}}- \lambda\overline{q^{1}} \\  24 k\overline{q^{7}}-4 \lambda \overline{q^{5}}+\frac{12}{\beta}\overline{q^{3}} & 18 k\overline{q^{6}}-3 \lambda\overline{q^{4}}+\frac{6}{\beta}\overline{q^{2}} & 12 k\overline{q^{5}}-2 \lambda\overline{q^{3}}+\frac{2}{\beta}\overline{q^{1}} &  6 k\overline{q^{4}}- \lambda\overline{q^{2}} \\ 24 k\overline{q^{8}}-4 \lambda \overline{q^{6}}+\frac{18}{\beta}\overline{q^{4}} & 18 k\overline{q^{7}}-3 \lambda\overline{q^{5}}+\frac{9}{\beta}\overline{q^{3}} & 12 k\overline{q^{6}}-2 \lambda\overline{q^{4}}+\frac{3}{\beta}\overline{q^{2}} & 6 k\overline{q^{5}}- \lambda\overline{q^{3}} \end{array}  \right) .  \label{seq:overarray1}
\end{align}
Here
\begin{align}    \overline{q^{n}}  \equiv \sqrt{\frac{2\pi}{4\beta \lambda}} \left[ q^{n}_{1}\mathrm{e}^{-\beta (kq^{4}_{1}-\lambda q^{2}_{1})}  + q^{n}_{2}\mathrm{e}^{-\beta (kq^{4}_{2}-\lambda q^{2}_{2})}  \right]   , \quad n=0,1,2,\cdots .\label{seq:oversaddle}\end{align}
Solving Eq.~(\ref{seq:overarray2}), we can derive that
\begin{equation} a_{1}^{*}=a_{3}^{*}=0, \quad a_{2}^{*}= -\frac{3\gamma}{8\lambda}, \quad a_{4}^{*}=\frac{\gamma k}{8\lambda^{2}}.              \label{seq:overbestpar}\end{equation}
Therefore, the best possible auxiliary potential takes the form
\begin{equation}  U_{a}(q,t)=\frac{\gamma \dot{\lambda}}{8\lambda^{2}}\left( kq^{4}-3\lambda q^{2}    \right).              \label{seq:overfinU1}\end{equation}
Note that the saddle-point approximation~(\ref{seq:overaction2}) applies when the distance between the maximum and the minimum of $ \beta U_{o}$, i.e., $A \equiv \beta\lambda^{2}/(4k)$, is much larger than 1.
Therefore, the auxiliary potential~(\ref{seq:overfinU1}) only works when the controlling parameter satisfies $\lambda(t) \gg \sqrt{4k/\beta}$, and it fails when $\lambda \to 0$.

Let us recall that, in the underdamped situation, the parameters $b_{2}^{*}$ and $b_{4}^{*}$ are still undetermined in the auxiliary potential~(\ref{eq:underdambestpoten}). Comparing Eq.~(\ref{seq:overfinU1}) with Eq.~(\ref{eq:underdambestpoten}), we find that both problems can be reconciled if assuming that the auxiliary potential takes the form
\begin{equation} U_{a}(q,p,t)=\frac{\beta \dot{\lambda}}{8\beta \lambda^{2}+12k} \left( 4\lambda qp+\gamma kq^{4}-3\gamma \lambda q^{2}   \right),          \label{seq:underdambestpoten2}\end{equation}
in the underdamped situation and
\begin{equation} U_{a}(q,t)=\frac{\gamma\beta \dot{\lambda}}{8\beta\lambda^{2}+12k}\left( kq^{4}-3\lambda q^{2}    \right),          \label{seq:overdambestpoten2}\end{equation}
in the overdamped situation. In this way, the parameters in Eq.~(\ref{eq:underdambestpoten}) take the forms $b_{2}^{*}=-3\gamma\beta \lambda/(8\beta\lambda^{2}+12k)$ and $b_{4}^{*}=\gamma\beta k/(8\beta\lambda^{2}+12k)$, which can be verified to satisfy the relation~(\ref{eq:pararelation}). Besides, the denominator in Eq.~(\ref{seq:overfinU1}) is amended to avoid divergence of the auxiliary potential in the limit $\lambda \to 0$. Note that Eq.~(\ref{seq:underdambestpoten2}) will reduce to Eq.~(\ref{seq:overdambestpoten2}) in the overdamped limit $m/\gamma \to 0$, which can support our assumptions about the form of the auxiliary potentials~(\ref{seq:underdambestpoten2}) and~(\ref{seq:overdambestpoten2}).

\section{Details of the simulation\label{SSec-seven}}
We simulate an underdamped Brownian particle moving in the double-well potential~(\ref{seq:Sunpotential}) and add the auxiliary potential~(\ref{seq:underdambestpoten2}) to approximately realize shortcuts to isothermality.
The motion of the Brownian particle is governed by the modified Langevin equation~(\ref{eq:underLangevinone}). There are two characteristic times $\tau_{p} \equiv m/\gamma$ and $\tau_{q} \equiv \gamma / \sqrt{k k_{B}T}$ in the system. Through introducing the characteristic length $l_{c} \equiv (k_{B}T/k)^{1/4}$, we can reduce the coordinate $\tilde{q} \equiv q / l_{c}$, the momentum $\tilde{p} \equiv p  \tau / ml_{c}$, the time $s \equiv t / \tau$, and the driving protocol $\tilde{\lambda} \equiv \lambda / (k l_{c}^{2})$. The modified Langevin equation~(\ref{eq:underLangevinone}) can be transformed into the dimensionless form:
\begin{align} \tilde{q}'=& \tilde{p} +  \alpha \tilde{\tau}^{2}\frac{\partial \tilde{U}_{a}}{\partial \tilde{p}}, \nonumber \\ \tilde{p}'= & -\alpha \tilde{\tau}^{2}\frac{\partial \tilde{U}_{o}}{\partial \tilde{q}}  - \alpha \tilde{\tau}^{2} \frac{\partial \tilde{U}_{a}}{\partial \tilde{q}}  - \tilde{\tau} \left( \tilde{p} +  \alpha \tilde{\tau}^{2}\frac{\partial \tilde{U}_{a}}{\partial \tilde{p}}  \right)+ \tilde{\tau}\sqrt{2\alpha\tilde{\tau}} \zeta(s), \label{seq:underdimensionlessLE}\end{align}
where $\tilde{\tau}\equiv \tau/\tau_{p}$ and $\alpha \equiv \tau_{p}/\tau_{q}$. The prime on a variable represents the derivative of that variable with respect to the time $s$. $\zeta(s)$ represents Gaussian white noise that satisfies $\langle \zeta(s) \rangle = 0$ and $\langle \zeta(s_{1}) \zeta(s_{2}) \rangle = \delta(s_{1}-s_{2})$.
The dimensionless form of the auxiliary potential takes
\begin{equation} \tilde{U}_{a}(\tilde{q},\tilde{p},s)=\frac{\tilde{\lambda}'}{\alpha\tilde{\tau}^{2}(8\tilde{\lambda}^{2}+12)}(4 \tilde{\lambda}\tilde{q}\tilde{p}+\tilde{\tau}\tilde{q}^{4}-3\tilde{\tau}\tilde{\lambda}\tilde{q}^{2}) . \label{seq:underdamdlessoveraux}\end{equation}
Equation~(\ref{seq:underdamdlessoveraux}) is solved by using the Euler algorithm
\begin{align}    \tilde{q}(s+\delta s) =& \tilde{q}(s) + \tilde{p}\delta s + \alpha \tilde{\tau}^{2}\frac{\partial \tilde{U}_{a}}{\partial \tilde{p}}\delta s, \nonumber \\  \tilde{p}(s+\delta s) =& \tilde{p}(s) -\alpha \tilde{\tau}^{2}\frac{\partial \tilde{U}_{o}}{\partial \tilde{q}} \delta s  - \alpha \tilde{\tau}^{2} \frac{\partial \tilde{U}_{a}}{\partial \tilde{q}}\delta s  - \tilde{\tau} \left( \tilde{p} +  \alpha \tilde{\tau}^{2}\frac{\partial \tilde{U}_{a}}{\partial \tilde{p}}  \right) \delta s + \tilde{\tau}\sqrt{2\alpha\tilde{\tau}\delta s} \theta (s), \label{seq:undereularalgor}\end{align}
where $\delta s $ is the time step and $\theta(s)$ is a random number sampled from Gaussian distribution with zero mean and unit variance.


\begin{thebibliography}{50}
\bibitem{Jarzynski1997}C. Jarzynski, \emph{Nonequilibrium Equality for Free Energy Differences}, Phys. Rev. Lett. \textbf{78}, 02690 (1997).

\bibitem{Zuckerman2002}D. M. Zuckerman and T. B. Woolf, \emph{Theory of a systematic computational error in free energy differences}, Phys. Rev. Lett. \textbf{89}, 180602 (2002).

\bibitem{Gore2003}J. Gore, F. Ritort, and C. Bustamante, \emph{Bias and error in estimates of equilibrium free-energy differences from nonequilibrium measurements}, Proc. Natl. Acad. Sci. U.S.A. \textbf{100}, 12564 (2003).

\bibitem{Jarzynski2006}C. Jarzynski, \emph{Rare events and the convergence of exponentially averaged work values}, Phys. Rev. E \textbf{73}, 046105 (2006).

\bibitem{Geng2017}G. Li, H. T. Quan, and Z. C. Tu, \emph{Shortcuts to isothermality and nonequilibrium work relations}, Phys. Rev. E \textbf{96}, 012144 (2017).

\bibitem{Albay2019}J. A. C. Albay, S. R. Wulaningrum, C. Kwon, P.-Y. Lai, and Y. Jun, \emph{Thermodynamic cost of a shortcuts-to-isothermal transport of a Brownian particle}, Phys. Rev. Research \textbf{1}, 033122 (2019).

\bibitem{Albay2020A}J. A. C. Albay, P.-Y. Lai, and Y. Jun, \emph{Realization of finite-rate isothermal compression and expansion using optical feedback trap}, Appl. Phys. Lett. \textbf{116}, 103706 (2020).

\bibitem{Albay2020N}J. A. C. Albay, C. Kwon, P.-Y. Lai, and Y. Jun, \emph{Work relation in instantaneous-equilibrium transition of forward and reverse processes}, New J. Phys. \textbf{22}, 123049 (2020).

\bibitem{Pancotti2019}N. Pancotti, M. Scandi, M. T. Mitchison, and M. Perarnau-Llobet, \emph{Speed-Ups to Isothermality: Enhanced Quantum Heat Engines through Control of the System-Bath Coupling}, Phys. Rev. X \textbf{10}, 031015 (2020).

\bibitem{Nakamura2020}K. Nakamura, J. Matrasulov, and Y. Izumida, \emph{Fast-forward approach to stochastic heat engine}, Phys. Rev. E \textbf{102}, 012129 (2020).

\bibitem{Plata2020}C. A. Plata, D. Gu\'{e}ry-Odelin, E. Trizac, and A. Prados, \emph{Building an irreversible Carnot-like heat engine with an overdamped harmonic oscillator}, J. Stat. Mech. (2020) 093207.

\bibitem{Iram2019}S. Iram, E. Dolson, J. Chiel, J. Pelesko, N. Krishnan, \"{O}. G\"{u}ng\"{o}r, B. Kuznets-Speck, S. Deffner, E. Ilker, J. G. Scott, and M. Hinczewski, \emph{Controlling the speed and trajectory of evolution with counterdiabatic driving}, Nat. Phys. \textbf{17}, 135 (2021).

\bibitem{Miller2000}M. A. Miller and W. P. Reinhardt, \emph{Efficient free energy calculations by variationally optimized metric scaling: Concepts and applications to the volume dependence of cluster free energies and to solid-solid phase transitions}, J. Chem. Phys. \textbf{113}, 7035 (2000).

\bibitem{Jarzynski2002}C. Jarzynski, \emph{Targeted free energy perturbation}, Phys. Rev. E \textbf{65}, 046114 (2002).

\bibitem{Vaikuntanathan2008}S. Vaikuntanathan and C. Jarzynski, \emph{Escorted Free Energy Simulations: Improving Covergence by Reducing Dissipation}, Phys. Rev. Lett. \textbf{100}, 190601 (2008).

\bibitem{Minh2011}D. D. Minh and S. Vaikuntanathan, \emph{Density-dependent analysis of nonequilibrium paths improves free energy estimates II. A Feynman-Kac formalism}, J. Chem. Phys. \textbf{134}, 034117 (2011).

\bibitem{Martinez2016NP}I. A. Mart\'{i}nez, A. Petrosyan, D. Gu\'{e}ry-Odelin, E. Trizac, and S. Ciliberto, \emph{Engineered swift equilibration of a Brownian particle}, Nat. Phys. \textbf{12}, 843 (2016).

\bibitem{Cunuder2016}A. Le Cunuder, I. A. Mart\'{i}nez, A. Petrosyan, D. Gu\'{e}ry-Odelin, E. Trizac, and S. Ciliberto, \emph{Fast equilibrium switch of a micro mechanical oscillator}, Appl. Phys. Lett. \textbf{109}, 113502 (2016).

\bibitem{Chupeau2018}M. Chupeau, S. Ciliberto, D. Gu\'{e}ry-Odelin, and E. Trizac, \emph{Engineered swift equilibration for Brownian objects: from underdamped to overdamped dynamics}, New. J. Phys. \textbf{20}, 075003 (2018).

\bibitem{Gauss1829}C. F. Gauss, \emph{\"{U}ber ein neues allgemeines Grundgesetz der Mechanik}, J. Reine Angew. Math. \textbf{4}, 232 (1829).

\bibitem{Evans1983}D. J. Evans, W. G. Hoover, B. H. Failor, B. Moran, and A. J. C. Ladd, \emph{Nonequilibrium molecular dynamics via Gauss's principle of least constraint}, Phys. Rev. A \textbf{28}, 1016 (1983).

\bibitem{Bright2005}J. N. Bright, D. J. Evans, and D. J. Searles, \emph{New observations regarding deterministic, time-reversible thermostats and Gauss's principle of least constraint}, J. Chem. Phys. \textbf{122}, 194106 (2005).

\bibitem{Sels2017}D. Sels and A. Polkovnikov, \emph{Minimizing irreversible losses in quantum systems by local counterdiabatic driving}, Proc. Natl. Acad. Sci. U.S.A. \textbf{114}, E3909 (2017).

\bibitem{Kolodrubetz2017}M. Kolodrubetz, D. Sels, P. Mehta, and A. Polkovnikov, \emph{Geometry and non-adiabatic response in quantum and classical systems}, Phys. Rep. \textbf{697}, 1 (2017).

\bibitem{Claeys2019}P. W. Claeys, M. Pandey, D. Sels, and A. Polkovnikov, \emph{Floquet-Engineering Counterdiabatic Protocols in Quantum Many-Body Systems}, Phys. Rev. Lett. \textbf{123}, 090602 (2019).

\bibitem{Butler2007}R. W. Butler, \emph{Saddlepoint Approximations with Applications}, (Cambridge, New York, 2007).

\bibitem{Guery2019}D. Gu\'{e}ry-Odelin, A. Ruschhaupt, A. Kiely, E. Torrontegui, S. Mart\'{i}nez-Garaot, and J. G. Muga, \emph{Shortcuts to adiabaticity: Concepts, methods, and applications}, Rev. Mod. Phys. \textbf{91}, 045001 (2019).

\bibitem{Saito1976}Y. Saito, \emph{Relaxation in a Bistable System}, J. Phys. Soc. Jpn. \textbf{44}, 388 (1976).

\bibitem{Hanggi1990}P. H\"{a}nggi, P. Talkner, and M. Borkovec, \emph{Reaction-rate theory: fifty years after Kramers}, Rev. Mod. Phys. \textbf{62}, 251 (1990).

\bibitem{Sun2003}S. X. Sun, \emph{Equilibrium free energies from path sampling of nonequilibrium trajectories}, J. Chem. Phys. \textbf{118}, 5759 (2003).

\bibitem{Coffey2004}W. T. Coffey, Yu. P. Kalmykov, and J. T. Waldron, \emph{The Langevin Equation}, 2nd ed. (World Scientific, New Jersey, 2004).

\bibitem{Mazonka1998}O. Mazonka, C. Jarzynski, and J. Bocki, \emph{Computing probabilities of very rare events for Langevin processes: a new method based on importance sampling}, Nucl. Phys. A \textbf{641}, 335 (1998).

\bibitem{Allen2005}R. J. Allen, P. B. Warren, and P. R. ten Wolde, \emph{Sampling rare switching events in Biochemical networks}, Phys. Rev. Lett. \textbf{94}, 018104 (2005).

\bibitem{Kundu2011}A. Kundu, S. Sabhapandit, and A. Dhar, \emph{Application of importance sampling to the computation of large deviations in nonequilibrium processes}, Phys. Rev. E \textbf{83}, 031119 (2011).


\bibitem{Funo2020}K. Funo, N. Lambert, F. Nori, and C. Flindt, \emph{Shortcuts to Adiabatic Pumping in Classical Stochastic Systems}, Phys. Rev. Lett. \textbf{124}, 150603 (2020).

\bibitem{Takahashi2020}K. Takahashi, K. Fujii, Y. Hino, and H. Hayakawa, \emph{Nonadiabatic Control of Geometric Pumping}, Phys. Rev. Lett. \textbf{124}, 150602 (2020).

\bibitem{Lu2017}Z. Lu and O. Raz, \emph{Nonequilibrium thermodynamics of the Markovian Mpemba effect and its inverse}, Proc. Natl. Acad. Sci. U.S.A. \textbf{114}, 5083 (2017).

\bibitem{Gal2020}A. Gal and O. Raz, \emph{Precooling Strategy Allows Exponentially Faster Heating}, Phys. Rev. Lett. \textbf{124}, 060602 (2020).

\bibitem{Kumar2020}A. Kumar and J. Bechhoefer, \emph{Exponentially faster cooling in a colloidal system}, Nature \textbf{584}, 64 (2020).

\bibitem{Deffner2020}S. Deffner and M. V. S. Bonan\c ca, \emph{Thermodynamic control-an old paradigm with new applications}, Europhys. Lett. \textbf{131}, 20001 (2020).

\bibitem{Reichl2009}Reichl L E, \emph{A Modern Course in Statistical Physics}, 3nd ed. (Wiley, Weinheim, 2009).

\bibitem{Masuda2010}S. Masuda and K. Nakamura, \emph{Fast-forward of adiabatic dynamics in quantum mechanics}, Proc. R. Soc. A. \textbf{466}, 1135 (2010).

\bibitem{Torrontegui2012}E. Torrontegui, S. Mart\'{i}nez-Garaot, A. Ruschhaupt, and J. G. Muga, \emph{Shortcuts to adiabaticity: Fast-forward approach}, Phys. Rev. A \textbf{86}, 013601 (2012).

\bibitem{Martinez2016}S. Mart\'{i}nez-Garaot, M. Palmero, J. G. Muga, and D. Gu\'{e}ry-Odelin, \emph{Fast driving between arbitrary states of a quantum particle by trap deformation}, Phys. Rev. A \textbf{94}, 063418 (2016).

\bibitem{Jarzynski2017}C. Jarzynski, S. Deffner, A. Patra, and Y. Suba\c{s}\i, \emph{Fast forward to the classical adiabatic invariant}, Phys. Rev. E \textbf{95}, 032122 (2017).

\bibitem{Patra2017}A. Patra and C. Jarzynski, \emph{Shortcuts to adiabaticity using flow field}, New. J. Phys. \textbf{19}, 125009 (2017).

\bibitem{Sekimoto2010}K. Sekimoto, \emph{Stochastic Energetics} (Springer, New York, 2010).


\end{thebibliography}
\end{document}